\documentclass[11pt,epsf]{article}

\usepackage{amsfonts}

\def\beq{\begin{eqnarray}}
\def\eeq{\end{eqnarray}}
\def\bea{\begin{eqnarray*}}
\def\eea{\end{eqnarray*}}

\def\centeron#1#2{{\setbox0=\hbox{#1}\setbox1=\hbox{#2}\ifdim
\wd1>\wd0\kern.5\wd1\kern-.5\wd0\fi
\copy0\kern-.5\wd0\kern-.5\wd1\copy1\ifdim\wd0>\wd1
\kern.5\wd0\kern-.5\wd1\fi}}
\def\ltap{\;\centeron{\raise.35ex\hbox{$<$}}{\lower.65ex\hbox{$\sim$}}\;}
\def\gtap{\;\centeron{\raise.35ex\hbox{$>$}}{\lower.65ex\hbox{$\sim$}}\;}

\def\singleandthirdspaced{\baselineskip=\normalbaselineskip\multiply
    \baselineskip by 130\divide\baselineskip by 100}

\newcommand{\newc}{\newcommand}
\newc{\qbar}{{\overline q}}
\newc{\Kahler}{K\"ahler }
\newc{\deltaGS}{\delta_{\rm GS}}

\newc{\bbar}[1]{\overline{#1}}

\begin{document}
\begin{titlepage}
  \begin{flushright}
    {\large hep-th/yymmnnn \\ SCIPP 09/12\\
    }
  \end{flushright}

  \vskip 1.2cm

  \begin{center}

    {\LARGE\bf Discrete $R$ Symmetries and Low Energy Supersymmetry}

    \vskip 1.4cm

    {\large Michael Dine and John Kehayias}
    \\
    \vskip 0.4cm {\it Santa Cruz Institute for Particle Physics and \\
      Department of Physics, University of California,
      Santa Cruz CA 95064  } \\
    \vskip 4pt

    \vskip 1.5cm

\begin{abstract}
  If nature exhibits low energy supersymmetry, discrete
  (non-$\mathbb{Z}_2$) $R$ symmetries may well play an important role.
  In this paper, we explore such symmetries.  We generalize gaugino
  condensation, constructing large classes of models which are
  classically scale invariant, and which spontaneously break discrete
  $R$ symmetries (but not supersymmetry).  The order parameters for
  the breaking include chiral singlets.  These simplify the construction
  of models with metastable dynamical supersymmetry breaking.  We
  explain that in gauge mediation, the problem of the cosmological
  constant makes ``retrofitting'' particularly natural -- almost
  imperative.  We describe new classes of models, with interesting
  scales for supersymmetry breaking, and which allow simple solutions
  of the $\mu$ problem.  We argue that models exhibiting such $R$
  symmetries can readily solve not only the problem of dimension four
  operators and proton decay, but also dimension five operators.  On
  the other hand, in theories of ``gravity mediation,'' the breaking
  of an $R$ symmetry is typically of order $M_p$; $R$ parity is
  required to suppress dimension four $B$ and $L$ violating operators,
  and dimension five operators remain problematic.
\end{abstract}

\end{center}

\vskip 1.0 cm

\end{titlepage}
\setcounter{footnote}{0} \setcounter{page}{2} \setcounter{section}{0}
\setcounter{subsection}{0} \setcounter{subsubsection}{0}

\singleandthirdspaced
\section{What Makes $R$ Symmetries Special}

While it has long been argued that theories which incorporate general
relativity cannot exhibit exact global continuous symmetries, discrete
symmetries are another matter.  When studying classical solutions of
critical string theories, one often encounters such symmetries on
submanifolds of the full moduli space of solutions \cite{gsw}; these
are believed, in general, to be discrete gauge transformations.
Within spaces of supersymmetric solutions, a particularly interesting
set of symmetries are the discrete $R$ symmetries.  These can often be
thought of as unbroken subgroups of the rotation group in higher
dimensions; in such cases, the fact that they transform the
supercharges is immediate.

We will reserve the term $R$ symmetry, in this paper, for symmetries
{\it other} than $\mathbb{Z}_2$ which rotate the supercharges.  Any
symmetry which multiplies all of the supercharges by $-1$ can be
redefined by adding a rotation by $2 \pi$, leaving an ordinary (non-R)
$\mathbb{Z}_2$.  Conventional $R$ parity in this sense, is not an $R$
symmetry.

There are several reasons to think that, if supersymmetry plays some
role in low energy physics, discrete $R$ symmetries might be relevant:
\begin{enumerate}
\item Perhaps the most important comes from the question of the
  cosmological constant (c.c.).  In order that the c.c.~be small, it
  is necessary that any constant in the superpotential be far smaller
  than $M_p^3$.  The only type of symmetry which can suppress such a
  constant is an $R$ symmetry.\footnote{Supersymmetric critical string
    theories have vanishing c.c.  classically, and the superpotential
    is protected in higher orders of perturbation theory by
    non-renormalization theorems.  In many cases, the vanishing of $W$
    appears an accident, from the point of view of the low energy
    theory (it is not accounted for by symmetries).  In flux vacua,
    vanishing of the cosmological constant is not typical, requiring
    $R$ symmetries, or accidental cancelations.  More generally, as
    Banks has repeatedly stressed, the only Minkowski space gravity
    theories of which we can claim any complete understanding exhibit
    supersymmetry and $R$ symmetries.  Similar remarks apply to
    would-be $\mu$ terms.}
\item In gauge-mediated models, supergravity effects should be
  unimportant for understanding SUSY breaking (we will make this
  statement more precise shortly), and the theorem of Nelson and
  Seiberg \cite{nelsonseiberg} requires a global $R$ symmetry in order
  to obtain supersymmetry breaking (in a generic fashion);
  correspondingly, an approximate global $R$ symmetry seems to be a
  requirement for metastable supersymmetry breaking \cite{iss}.
  Discrete $R$ symmetries are a particularly plausible way in which to
  account for such approximate continuous symmetries.
\item Discrete symmetries have long been considered in supersymmetric
  theories to suppress proton decay \cite{protondecay} and $R$
  symmetries might be particularly effective in suppressing dimension
  four -- and five -- operators which violate $B$ and $L$
  \cite{japaneser}.
\end{enumerate}

In supersymmetric, $R$ symmetric theories, because $W$ transforms
under the symmetry, there is a close tie between the scale of
supersymmetry breaking and the scale of $R$ breaking, if the c.c.,
$\Lambda$, is small.  This is because \beq \vert \langle W \rangle
\vert \sim \vert \langle F \rangle \vert M_p.
\label{wvev}
\eeq We will see in this paper that in ``gravity mediated" theories
($\sqrt{F} \sim 10^{11}$ GeV), except under special circumstances,
this implies that the $R$ symmetry is broken by scalar field vevs of
order $M_p$.  As a result, the $R$ symmetry does not significantly
constrain the low energy Lagrangian, and cannot account for proton
stability or other phenomena.  On the other hand, in gauge mediation,
the contribution of the SUSY breaking interactions to $W$ is typically
much smaller than eqn.~\ref{wvev}.\footnote{This point was first
stressed to one of the authors many years ago by T. Banks.}  New
interactions are required; the needed scale is precisely that which
enters in retrofitted models \cite{retrofitted}.  This suggests that
retrofitting, rather than being a sort of Rube Goldberg contraption to
implement (metastable) supersymmetry breaking, may be essential to
understanding the smallness of the
c.c. 


These types of considerations lead us to consider more broadly a set
of questions about discrete $R$ symmetries.
\begin{enumerate}
\item Spontaneous breaking of discrete $R$ symmetries is familiar in
  pure supersymmetric gauge theories (gaugino condensation), and in
  theories with massive matter fields.  In section
  \ref{discretedynamics}, we discuss a set of theories which
  generalize gaugino condensation.  These models are quite close to a
  set of theories considered some time ago by Yanagida
  \cite{yanagida}.  This larger set of theories includes examples with
  gauge singlets and matter fields.  These theories, like the pure
  gauge theory, are classically scale invariant, and exhibit intricate
  discrete symmetries.
\item In section \ref{gravitymediation}, we explain why discrete $R$
  symmetries are typically badly broken in gravity-mediated models.
\item In the framework of gauge mediation, discrete $R$ symmetries are
  distinctly more interesting.  In section \ref{gaugemediation}, we
  explain why, in gauge mediation, retrofitting is almost
  inevitable if one is to understand the smallness of the cosmological
  constant.\footnote{Banks has put forth an alternative proposal, with
    the framework of {\it Cosmological Supersymmetry Breaking}
    \cite{csb}, to understand the relation of these scales.  In his
    proposal, discrete $R$ symmetries also play an important role,
    both in understanding the c.c.~and in suppressing dangerous rare
    processes.}  We construct simple, gauge-mediated models in this
  framework.
\item In section \ref{mutermsection}, we revisit an earlier suggestion
  of Yanagida's for solving the $\mu$ problem of gauge-mediated models, noting that (for
  similar reasons as in \cite{yanagida}) it is readily understood in
  retrofitted models with singlets, and argue that a large value of
  $\tan \beta$ is typical.  (In gravity mediation, a source for $\mu$ with
  a similar flavor was proposed in \cite{nillesmu}.  In gauge mediation, the issue is more severe,
  since $B_\mu$ tends to be parametrically too large; our solution is particularly
  directed at this issue.)
\item While there have been studies of discrete $R$ symmetries and
  proton decay, almost all of these are within the context of
  intermediate scale supersymmetry breaking.  While we argue that such
  symmetries are unlikely to play a significant role in gravity
  mediated models, in the context of gauge mediated models, discrete
  $R$ symmetries {\it can} be effective in suppressing dimension four
  and five contributions to proton decay.  We explain in section
  \ref{protondecay} why one cannot apply anomaly constraints for this
  question, and give simple examples of symmetries which suppress some
  or all dangerous dimension four or dimension five operators.
\item Discrete $R$ symmetries might have cosmological relevance.  They
  have been considered for models of inflation
  (e.g.~\cite{japaneseinflation}) and might well play a role in AD
  baryogenesis \cite{drt}.  Discrete $R$ symmetries can lead to the
  suppression of higher dimension operators and thus account for very
  flat directions which might be relevant for these phenomena.  The
  observations of this paper could well be relevant to to
  understanding Peccei-Quinn symmetries as well.  We comment on these
  issues in the concluding section, leaving a more thorough
  investigation for future work.
\end{enumerate}

\section{Generalizations of Gaugino Condensation}
\label{discretedynamics}

Gaugino condensation is considered in many contexts, but its principal
distinguishing feature is that it breaks a discrete $R$ symmetry
without breaking supersymmetry.  Many other models, such as
supersymmetric QCD with massive quarks, dynamically break such
symmetries, but in thinking about a variety of questions, it would be
helpful to have models like pure susy gauge theory, with scales
generated by dimensional transmutation.  In this section, we develop a
class of such models.

As an example, consider an $SU(N)$ gauge theory with $N_f < N$
massless flavors, and, in addition, $N_f^2$ singlets,
$S_{f,f^\prime}$\footnote{This model has been considered previously in
  \cite{yanagida}, who find the solutions obtained in
  eqn.~\ref{discretesolutions}} .  Include a superpotential:
\beq
W = y S_{f,f^\prime} \bbar Q_{f^\prime} Q_f -{1\over 3} \gamma {\rm Tr}~S^3
\eeq
where the last term denotes a general cubic coupling of the
$S_{f,f^\prime}$ fields. For convenience, we have taken the
superpotential to respect an $SU(N_f)$ symmetry; $\gamma$ and $y$ can
be taken real, by field redefinitions. This model possesses a discrete
$\mathbb{Z}_{2(3N-N_f)}$ $R$ symmetry, which is free of anomalies
(this remains true away from the $SU(N_f)$ symmetric limit). Calling
\beq
\alpha = e^{2 \pi i \over 6N-2N_f}
\eeq
we can take the transformation laws of the various fields to be:
\beq
\lambda \rightarrow \alpha^{3/2} \lambda~~~~~S_{f,f^\prime}
\rightarrow \alpha S_{f,f^\prime}~~~~ (Q,\bbar Q) \rightarrow \alpha(Q,\bbar Q).
\eeq

For $N_f < N$, treating $\gamma$ and $y$ as small, we can analyze the
system by including the familiar non-perturbative superpotential of
$SU(N)$ QCD with $N_f$ flavors \cite{ads}
\beq
W_{dyn} = (N-N_f) {\Lambda^{3N-N_f \over N-N_f} \det(\bbar Q Q)^{-{1 \over N-N_f}}}.
\eeq
In the $SU(N_f)$ symmetric limit, the ${\partial W \over \partial Q}, {\partial W \over \partial S} =0$ equations admit solutions of the form
\beq
S_{f,f^\prime} = s \delta_{f,f^\prime}~~~~ Q_f \bbar Q_{f^\prime} = v^2
\delta_{f,f^\prime}.
\eeq
with
\beq
v^2 = \left( {\gamma \over y^3}
\right )^{N-N_f \over 3N - N_f} \alpha^{2k} \Lambda;
~~~~s = \left({y^{N_f} \over \gamma^N} \right )^{1 \over 3N - N_f} \alpha^{2k}\Lambda.
\label{discretesolutions}
\eeq
Perturbing away from the symmetric limit, one can then check that
there is no qualitative change in the solutions (e.g.~the number is
unchanged).  For $N_f \ge N$, the theory has baryonic flat directions,
and does not have a discrete set of supersymmetric ground states.
Adding additional singlets and suitable (non-renormalizable)
couplings, one can again spontaneously break the discrete symmetries.
One can also consider generalizations to other gauge groups and to
different matter content.  These and related matters will be
thoroughly explored in reference \cite{kehayias}.

\section{Discrete $R$ Symmetries in Gravity Mediation}
\label{gravitymediation}

Supergravity models, without $R$ symmetries,
seem to have a virtue with respect to the cosmological constant.
If one has a pseudo-modulus, $Z$, with, say, a constant superpotential, $W_0$,
then if the field $Z$ varies over the Planck scale, the positive and negative
terms in the potential are of a similar order of magnitude.  This is different
than the case of gauge mediation, where additional interactions, or a rather peculiar
value for $W_0$, are required to account for a small cosmological constant.

In a hidden sector supergravity model, with an underlying $R$
symmetry, one might hope to break the $R$ symmetry through hidden
sector dynamics.  More precisely, one might hope that the same
dynamics which is responsible for supersymmetry breaking would
spontaneously break the $R$ symmetry, generating a superpotential
of a size suitable to cancel the cosmological constant.

However, in such theories, quite generally, the $R$
symmetry breaking is large (broken by Planck-scale vevs).  The
difficulty is associated with the cosmological constant.  In
supergravity theories, one requires that $W \sim F M_p$.  In a theory
with an $R$ symmetry at the scale of the hidden sector dynamics, a
constant in the superpotential is forbidden, so one might expect that
$W \sim F Z$, for some field, $Z$, transforming non-trivially under
the $R$ symmetry.  Then one would have $Z \sim M_p$.

One can almost make this rigorous.  Consider, first, O'Raifeartaigh
(OR) type models, with a continuous $R$ symmetry.  Assuming that all
fields are small compared to $M_p$, we can limit our considerations to
renormalizable theories.  In this case, one can show that, at tree
level, there is a chiral field, $Z$, whose fermionic component is the
Goldstino, and whose scalar components are a (tree level) modulus and
a massless pseudoscalar \cite{ray, ks}.  The effective superpotential
of the theory is:
\beq
W = Z F.
\label{guido}
\eeq
One can readily prove, in such models, a bound \cite{dfk}:
\beq
\vert \langle W \rangle \vert \le {1 \over 2} f_r \vert F\vert
\label{dfkinequality}
\eeq
where $f_r$ is the r-axion decay constant.  In order to cancel the
cosmological constant, one needs $f_r \sim M_p$.
The inequality, eqn.~\ref{dfkinequality}, appears rather general \cite{dfk}.

If supersymmetry is dynamically broken in the hidden sector in a
theory without flat directions, there are no large field vevs, and
the c.c.~is large and positive.  In theories with metastable breaking,
one typically has, at most, approximate moduli, so again the c.c.~is
positive.

One can attempt to require that the $R$ symmetry is broken at some
scale higher than the intermediate scale, by some other dynamics.  For
example, there could be additional gauge interactions at a scale
$\Lambda^3 = F M_p$\cite{dgt}, as might be expected in retrofitted models.  But
the hidden sector necessarily possesses a field(s), $Z$, neutral under
the $R$ symmetry, to generate the needed $F$ term; it couples to
$W_\alpha^2$ through a term of the form $ W_Z = ZW_\alpha^2/M_p$.
However, because $Z$ is neutral, the coupling
\beq
{Z^2 \over M_p^2}W_\alpha^2
\eeq
is also allowed, generating a tadpole for $Z$, and
yielding an expectation value of order $M_p$.

More intricate constructions, for example using singlet fields as in
the models of section \ref{discretedynamics}, and introducing
additional symmetries, might achieve the desired structure, but the
required models are clearly complicated.  We conclude from this that
theories with gravity-mediated supersymmetry breaking will typically
break $R$ symmetries by large amounts.  Such symmetries will not
suppress proton decay, or play other interesting roles in low energy
dynamics.

A particularly interesting proposal to understand a small $W$ in theories with discrete (and approximate
continuous) $R$ symmetries has been put forward in \cite{nillesr}.  These authors generate a small
constant in the superpotential in theories with a Fayet-Iliopoulos term, in heterotic orbifold
models.  This provides a small parameter.
In the low energy theory, an expectation value for $W$ arises at very high order in this small
parameter as a consequence of discrete symmetries.  As in our previous examples,
this corresponds to the breaking of the $R$ symmetry at energy scales well above
the scale of supersymmetry breaking.

\section{Discrete $R$ Symmetries and Retrofitted Models}
\label{gaugemediation}

Having established that $R$ symmetries are not of interest in gravity
mediation, our focus will be on theories with dynamical supersymmetry
breaking and gauge mediation.  One simple class of gauge-mediated
models begins with retrofitted O'Raifeartaigh (OR) models
\cite{retrofitted}.  At first sight, these models seem rather
contrived, with singlets and additional gauge interactions added just
so.  But when one thinks about the cosmological constant, retrofitting
takes on an aspect of inevitability.  One of the disturbing features
of gauge mediated models, as opposed to models with supersymmetry
broken at an intermediate scale, is the need for dynamics at some much
larger scale to account for the vanishing of the c.c.~(or,
alternatively, for a constant in the superpotential with some very
large scale, unrelated to any other scale seemingly required for the
models).  Retrofitted models offer a possible solution to this puzzle.
Consider the simplest of retrofitted models:
\beq
W = {c \over 32\pi^2}{Z~W_\alpha^2 \over M_p} + Z A^2 + M YA.
\label{basicretro}
\eeq
Calling
\beq
\langle \lambda \lambda \rangle = N\Lambda^3 e^{-{ cZ \over N M_p}}
\approx W_0 - \mu^2 Z ~~~~\mu^2 ={c \Lambda^3 \over M_p}~~~~W_0 = {N \Lambda^3},
\eeq
the low energy effective superpotential is (for $Z \ll M_p$):
\beq
W=W_0 + Z (A^2 - \mu^2) + Z
A^2 + M YA,
\eeq
a simple OR model, with a constant of suitable order
of magnitude to give a small c.c.  We see that $\langle W \rangle$ is
automatically of the correct order of magnitude to cancel the
c.c.\footnote{While the orders of magnitude are correct, it is still
  necessary to tune the constants in the lagrangian to obtain the high
  degree of cancelation required in nature.  As written, this can be
  achieved by tuning the constant $c$.  Whether this might be
  explained through Weinberg's argument \cite{weinberg}, or in some
  other way, we will not address here.}  The goldstino decay constant,
$F \approx \mu^2$.  In this model, one would still like to account for
the scale $M$.  One can consider the possibilities that $M \gg \mu$,
or $M \sim \mu$.  The scale relevant to low energy soft breaking terms
is \beq \Lambda_m^2 = {\vert F \vert^2 \over M^2} \eeq so the case of
$M \sim \mu$ corresponds to relatively low scale breaking (say $10^5$
GeV), while $M \gg \mu$ corresponds to high scale gauge mediation.  In
ref.~\cite{dm}, an alternative scaling was suggested: \beq \mu^2 \sim
{W_\alpha^4 \over M_p^4} \sim M^2 \eeq but the critical relation
required for canceling the cosmological constant, $W \sim F M_p$, is
spoiled; the scale of the new interactions is much too large.  In the
next subsection, we write models with a hierarchy between $M^2$ and
$F$.  In subsection \ref{single}, we develop models with $M^2 \sim F$,
exploiting the models of section \ref{discretedynamics} to avoid the
problematic relations of ref \cite{dm}.

\subsection{Models with a Hierarchy of Scales}
\label{hierarchymodel}

The ability to construct models with gauge singlet fields which break
discrete $R$ symmetries opens up a wide range of model-building
possibilities.  We have stressed that the cosmological constant points
in the direction of retrofitted OR, i.e.~OR models in which the
dimensionful parameters of the theory are determined by some high
scale dynamics.  Model building with OR models is challenging in and
of itself.  Shih \cite{shihr} studied O'Raifeartaigh models quite
generally, showing that in theories with fields with only (continuous)
$R$ charges $0$ or $2$, the $R$ symmetry is not spontaneously broken.
He also exhibited simple models which violate this condition, and {\it
  do} break the $R$ symmetry.  Perhaps the simplest example is
provided by a theory with fields $\phi_{\pm 1},\phi_{3},X_2$, where
the subscripts denote the $R$ charge, and with superpotential:
\beq
W = X_2 (\phi_1 \phi_{-1} - \mu^2) + M_1 \phi_1 \phi_1 + M_2 \phi_{-1}\phi_3.
\label{shihw}
\eeq
Motivated by this model, consider a theory with fields
$X_0,S_{2/3},\phi_{0}, \phi_{2/3}, \phi_{4/3}$, where the subscript
denotes the discrete $R$ charge ($\phi_q \rightarrow \alpha^{q}
\phi_q$, where $\alpha$ is some appropriate root of unity).  $S_{2/3}$
is a field with a large mass and an $R$ symmetry breaking vev,
presumed to arise from some high scale dynamics as in the models of
section \ref{discretedynamics}.  The superpotential is
\beq
W = {1 \over M_p} X_0 S_{2/3}^3
+ y X_0 \phi_{2/3} \phi_{4/3} + \lambda_1 S_{2/3} \phi_{2/3}
\phi_{2/3} + \lambda_2 S_{2/3} \phi_{4/3} \phi_{0}
\eeq
(up to terms involving higher dimension operators).\footnote{In order that this
  superpotential be the most general consistent with symmetries, it is
  necessary, beyond the $R$ symmetry, to impose a $\mathbb{Z}_2$ under
  which the $\phi$'s are even and the other fields are odd, and an
  additional symmetry to forbid a mass term for
  $\phi_{4/3}\phi_{2/3}$.  Without a mass term, the model possesses a
  $U(1)$ symmetry, under which the fields have charges:
$$X:  -3;  S:  1; \phi_{2/3}:   -1/2;  \phi_{4/3}:  7/2;  \phi_0:   -9/2.$$
A discrete subgroup is enough to eliminate the unwanted term.}  The
expectation value of $S$ leads to mass terms for the $\phi$ fields;
the resulting low energy effective theory is that of eqn.~\ref{shihw},
with
\beq
M_1 = \lambda_1 S_{2/3}~~~M_2 = \lambda_2 S_{2/3}~~~\mu^2 =
-{S_{2/3}^3 \over M_p}.
\eeq
Below the scale $S_{2/3}$, the theory possesses an accidental,
(approximate) continuous $R$ symmetry which is spontaneously broken.

Note that the mass terms are large compared to the effective $F$ term,
$M^2 \gg \mu^2$.  Assuming $X$ couples to some messenger fields, with
coupling $W_{mess} = X \bbar M M$, the scale of the low energy soft
terms is set by $F/M$
\beq
\Lambda_m = {S^2 \over M_p}.
\label{lambdam}
\eeq
If $\Lambda_m \sim 10^{5}$ GeV, for example, then we have $S \sim
10^{11.5}$ GeV.

Messengers are readily coupled to this system.  For example, a model
with minimal gauge mediation is obtained by coupling $X_0$ to
messengers $M$ and $\bbar M$ transforming as a $5$ and $\bbar 5$ of
$SU(5)$.  Models of general gauge mediation require more intricate
hidden sectors.

\subsection{O'Raifeartaigh Models with a Single Dimensionful
  Parameter}
\label{single}

The simplest O'Raifeartaigh model \beq W = X (A^2 -\mu^2) + m Y A \eeq
has two dimensionful parameters, one with dimensions of mass and one
with dimensions of mass-squared.  In order to obtain gauge mediation
with a low scale of supersymmetry breaking, one wants $m^2 \sim
\mu^2$.  On the other hand, if the underlying scale is $M_p$, and if
both parameters arise from gaugino condensation in some new gauge
group, then \beq \mu^2 \sim {\Lambda^3/M_p}~~~ m \sim{ \Lambda^3 \over
  M_p^2} \eeq i.e.  \beq m^2 \sim \mu^2 {\mu^2 \over M_p^2} \eeq To
avoid this, in \cite{dm} it was suggested that $\mu^2$ might arise
from a coupling to $W_\alpha^4/M_p^4$.  However, this is not
compatible with our suggested explanation of the cancelation of the
c.c.; the scale of $W_\alpha^2$ is much too large.  In this section,
we consider an alternative type of O'Raifeartaigh model, in which the
only dimensionful parameter has dimensions of mass-squared.

To illustrate some of the issues, consider, first, a model with three
fields: \beq W = X (A^2 - \mu^2) + \lambda Y A^2.  \eeq This model
formally breaks supersymmetry, but it is not attractive as one linear
combination of $X$ and $Y$ decouples from $A$.  So instead consider a
model with additional fields: \beq W =y X (AB - \mu^2) +\lambda_1 Y
A^2 + \lambda_2 ZB^2 \eeq This model breaks supersymmetry, and none of
the fields decouple.  The model, however, is not the most general
consistent with symmetries, and the Coleman-Weinberg calculation leads
to an unbroken $R$ symmetry.

We can avoid these difficulties, by again taking Shih's model, and
replacing both mass terms with a field $\Phi_0$:
\beq
W = X (\phi_{-1} \phi_1 - \mu_X^2) + \lambda_1 \Phi_0 \phi_1 \phi_1 + \lambda_2 \Phi_0 \phi_{-1} \phi_3 + \chi(\Phi_0^2 - \mu_\chi^2) + \epsilon X \Phi_0^2.
\eeq
In order that this model be the most general consistent with
symmetries, in addition to the $R$ symmetry we impose a $\mathbb{Z}_4$
symmetry under which
\beq
\Phi_0 \rightarrow -\Phi_0;
\phi_1 \rightarrow i \phi_1;
\phi_{-1} \rightarrow -i \phi_{-1};
\phi_3 \rightarrow -i \phi_3.
\eeq
We have {\it defined} $X$ to be the field which couples to $\phi_1 \phi_{-1}$.

If we first set $\epsilon=0$, and take $\mu_\chi^2 \gg \mu_X^2$, the
model is quite simple to analyze.  The fields $\Phi_0$ and $\chi$ are
massive, and integrating them out yields the model of
eqn.~\ref{shihw}.  There is, as in that model, a flat direction in the
theory, which, for a range of parameters, is stabilized with a
non-vanishing vev for $X$ (breaking the approximate, continuous $R$
symmetry of the low energy theory).  Turning on a small, non-zero
$\epsilon$ does not yield qualitative changes in the theory.  The only
light field, for $\epsilon = 0$, is the pseudomodulus $X$.  For
small $\epsilon$, there is still a pseudomodulus.  The equations
${\partial W \over \partial \Phi} =0$ still have a one (complex)
parameter set of solutions; the pseduomodulus is now a linear
combination of $\chi$ and $X$.  Similarly, there is no qualitative
change if $\mu_X^2$ is comparable to, but slightly less than,
$\mu_\chi^2$.  Both of these parameters can arise through retrofitting
as in \ref{basicretro}, and satisfy the critical relation.

\section{$R$ Symmetries and the $\mu$ Problem of Gauge Mediation}
\label{mutermsection}

Retrofitting has been discussed as a solution to the $\mu$ problem
\cite{yanagida, thomas,green}.  If the source of the $\mu$ term is a
coupling of the gaugino condensate responsible for the hidden sector
$F$ term, \beq W_{\mu} = {W_\alpha^2 \over M_p^2} H_U H_D \eeq the
resulting $\mu$ term is very small; it would seem necessary to
introduce still {\it another} interaction, with a higher scale.  Not
only does this seem implausibly complicated, but it is once more
problematic from the perspective of the c.c.  Models with singlets, on
the other hand, allow lower dimension couplings and larger $\mu$ terms
\cite{yanagida}.  For example, a model such as that of section
\ref{hierarchymodel} provides an interesting possible solution to the
$\mu$ problem.  If the product $H_U H_D$ has $R$ charge $2/3$, it can
couple to $S^2/M_p$ with coupling $\lambda$.  This gives a $\mu$ term
whose order of magnitude is $\lambda \Lambda_m$ ($\Lambda_m$ is the
low energy mass scale of eqn.~\ref{lambdam}), \beq W_\mu = \lambda
{S_{2/3}^2 \over M_p} H_U H_D.
\label{muterm}
\eeq The $F$ component of $S$ is naturally of order $m_{3/2}^2$, so
this does not generate an appreciable $B_\mu$ term; the $B_\mu$ term
must be generated at one loop, or through the operator \beq W_{B_\mu}
= {1 \over M_p^2} X_0 S_{2/3}^2 H_U H_D \eeq which can lead to an
appreciable $B_\mu$ term if $S \sim 10^{12}$ GeV.  For $S < 10^{12}$,
renormalization group evolution generates $B_\mu$.  A rough
calculation yields $\tan \beta \sim 30$.  Alternative structures lead
to different scaling relations; these, as well as a more detailed
analysis of $\tan \beta$, will appear elsewhere \cite{dmmu}.

We can similarly solve the $\mu$ problem in the single-scale models of
section \ref{single}.  Now, if ${\Lambda^3\over M_p} \sim S^2 \sim 10^{13}$ GeV$^2$, say, then
\beq
W_\mu = {S^2 \over M_p}H_U H_D
\eeq
yields a $\mu$ term of a suitable size.

\section{Discrete $R$ Symmetries and Proton Decay}
\label{protondecay}

Most supersymmetric model building seeks to suppress dangerous
dimension four lepton and baryon number violating operators by
imposing $R$ parity.  We have remarked that $R$ parity is not really
an $R$ symmetry at all.  Unlike the $R$ symmetries we are focussing on
in this paper, there is no requirement that it be broken; this leads,
most strikingly, to stable dark matter.  While discrete $R$ symmetries
might forbid dangerous dimension four {\it and} dimension five
operators, these symmetries must be broken; the size of this breaking,
and the transformation properties of the fields, will control the size
of $B$ and $L$ violating effects \cite{japaneser}.

In model building with discrete symmetries, one would seem to have a
great deal of freedom in both the choice of symmetry group and in the
transformation properties of the fields.
Many authors, in attempting to use discrete symmetries ($R$ or
non-$R$) to forbid proton decay, impose a variety of constraints.  For
example, the authors of \cite{japaneser}, who focus, as we do, mainly
on discrete $R$ symmetries, require:
\begin{enumerate}
\item Absence of anomalies.
\item $\mu$ term forbidden in the superpotential.
\item Kahler potential terms permitted which give rise to a $\mu$ term
  of order the supersymmetry breaking scale.
\end{enumerate}

The anomaly constraints cannot be imposed, however, without making
very strong assumptions about the microscopic theory.  Any $R$
symmetry must be spontaneously broken, by a substantial amount, in
order to account for the (near) vanishing of the c.c.  From the
perspective of anomaly cancelation, this means that there may be
massive states, at the TeV scale of higher, which contribute to
anomalies.  Such couplings can also generate the $\mu$ term.  So, in
fact, one has few ways of constraining the microscopic theory from low
energy considerations.

Most discussions of the use of $R$ symmetries to suppress proton decay
are framed in the context of gravity mediation, and we have seen that
once one requires a small cosmological constant, this is problematic.
So our focus here will be on gauge mediated models.

\subsection{$R$ Symmetries in Gauge Mediation}

It is easy to see that discrete symmetries can suppress all unwanted
dimension four and five operators.  To illustrate this point, suppose
that the theory possesses {\it conventional $R$ parity}, in addition
to an $R$ symmetry, under which all quark and lepton superfields are
neutral, while the Higgs transform like the superpotential\footnote{This particular
assignment forbids dimension five operators which would generate a Majorana
neutrino mass.  Different choices can achieve this.  Model building of this type
can be restricted in interesting ways by requiring unification or cancelation of anomalies,
but neither of these are required by general principles \cite{banksdinediscrete}.}.  This
forbids all dangerous dimension four and dimension five operators.
Once $R$ symmetry breaking is accounted for, dimension five operators
may be generated, but they will be highly suppressed.

We can contemplate more interesting symmetries, which do not include
$R$ parity, and for which the Higgs, quarks and leptons have more
intricate assignments under the $R$ symmetry.  In the absence of $R$
parity, given that the $R$ symmetry is necessarily broken, dangerous
dimension four operators will be generated, and it is important that
they be adequately suppressed.  Consider, first, the case where the
$R$ symmetry is broken by a gaugino condensate in a pure gauge theory.
Suppose that $B$ and $L$-violating operators of the form
\beq
{\delta W}_{b,l} \sim {W_{\alpha}^2 \over M_p^3} \Phi \Phi \Phi
\eeq
are permitted by the symmetries.  Even if $\sqrt{F}$ is as large as
$10^{9}$ GeV, $W_{\alpha}^2/M_p^3 \approx 10^{-18}$, more than adequately
suppressing proton decay.

In the presence of a singlet field such as $S$, the constraints are
more severe, however.  Even in the low gravitino mass case, the small
parameter, $S/M_p$, is of order $10^{-9}$.  So suppression of
dangerous operators by a single factor of $S$ is not adequate.  One
requires that many operators be suppressed by two powers of $S$.


\section{Conclusions}
\label{conclusions}

From this discussion, a coherent framework for electroweak symmetry
breaking due to supersymmetry appears to emerge.  Many of the
problematic aspects of most susy model building are resolved in
retrofitted models with $R$ symmetries.  The elements of the framework
are:
\begin{enumerate}
\item Gauge mediation: As in all gauge mediated models, the problem of
  flavor changing neutral currents is eliminated.
\item Discrete $R$ symmetries, spontaneously broken: such symmetries
  are likely to play a role in gauge-mediated models, e.g.~to account
  for the approximate $R$ symmetries needed for metastable
  supersymmetry breaking.  We have exhibited large classes of models
  which generalize gaugino condensation, in that they break discrete
  $R$ symmetries without breaking supersymmetry.
\item Metastable supersymmetry breaking: As illustrated by the ISS and
  retrofitted models, metastable supersymmetry breaking provides a
  rich setting in which to obtain dynamical supersymmetry breaking.
\item Retrofitting: In addition to providing a very simple realization
  of metastable supersymmetry breaking, retrofitting resolves the
  puzzle endemic to gauge-mediated models of the mismatch of scales
  required to cancel the cosmological constant.  The most troubling
  feature of the retrofitted models -- additional interactions
  introduced solely to account for the OR scale -- automatically
  yields a term in the superpotential of the correct order of
  magnitude.
\item $\mu$ term: This is easily generated in this framework,
  without elaborate additional sets of fields and/or arbitrary scales.
  At the messenger scale, $B_\mu$ vanishes; renormalization group
  evolution generates $B_\mu$ and large (but not excessively large)
  $\tan \beta$.
\item The problem of dimension five operators and proton decay can be
  readily resolved in models with $R$ symmetries, {\it provided that
    the scale of supersymmetry breaking is low}, as in gauge
  mediation.
\end{enumerate}

There are many open questions in this framework, some of which we have
indicated.  The space of models which dynamically break discrete $R$
symmetries -- the generalization of gaugino condensation -- needs
further exploration.  More detailed analysis of the $B_\mu$ term, and
its implications for $\tan \beta$, is warranted.  Another area where
discrete $R$ symmetries might play an important role is cosmology.
Such symmetries can lead to flat or nearly flat directions, which
might be important in inflation and/or baryogenesis.  These questions
will all be taken up in future work.

There are also problems we have not dealt with here, most notably the
so-called little hierarchy.  We have little to add on this question.
The little hierarchy might be ameliorated by ``squashing" of the
spectrum, as is possible in models of General Gauge Mediation
\cite{ggm}.  It is also possible that some sort of mild anthropic
constraint might account for a small hierarchy.  For example, in a
given model of inflation and reheating, there will be constraints on
the mass of the gravitino.  This, in turn, might force a higher scale
of supersymmetry breaking than naive tuning arguments would suggest.
Another class of questions has to do with axions.  These, too, might
force a rather high scale of supersymmetry breaking \cite{cdfu}.

\noindent
{\bf Acknowledgements:}
We thank Guido Festuccia, Zohar Komargodski, John Mason, Nathan
Seiberg, and David Shih for conversations and careful readings of an
early version of the manuscript.  This work was supported in part by
the U.S. Department of Energy.

\end{document}